\documentclass[twocolumn,showpacs,preprintnumbers,amsmath,amssymb]{revtex4}

\usepackage{graphicx}
\usepackage{dcolumn}
\usepackage{bm}

\begin{document}

\preprint{}

\title{ Gap Function with Point Nodes in Borocarbide Superconductor YNi$_2$B$_2$C}

\author{K.~Izawa$^1$, K.~Kamata$^1$, Y.~Nakajima$^1$, Y.~Matsuda$^1$, T.~Watanabe$^2$, M.~Nohara$^3$, H.~Takagi$^3$, P.~Thalmeier$^4$, and K.~Maki$^5$}

\affiliation{$^1$Institute for Solid State Physics, University of Tokyo, Kashiwanoha 5-1-5, Kashiwa, Chiba 277-8581, Japan}%
\affiliation{$^2$Department of Applied Chemistry, University of Tokyo, Hongo 7-3-1, Bunkyo-ku, Tokyo 113-8656, Japan}%
\affiliation{$^3$Department of Advanced Materials Science, University of Tokyo,  Hongo 7-3-1, Bunkyo-ku, Tokyo 113-0033, Japan}
\affiliation{$^4$Max-Planck-Institute for the Chemical Physics of Solid, N\"othnitzer Str.40, 01187 Dresden, Germany}%
\affiliation{$^5$Max-Planck-Institute for the Physics of Complex Systems, N\"othnitzer Str.38, 01187 Dresden, Germany}%


\begin{abstract}

	To determine the superconducting gap function of a borocarbide superconductor YNi$_2$B$_2$C,  the $c$-axis thermal conductivity $\kappa_{zz}$ was measured in a magnetic field  rotated in various directions relative to the crystal axes.   The angular variation of $\kappa_{zz}$ in {\boldmath $H$} rotated within the $ab$-plane shows a peculiar fourfold oscillation with narrow cusps.   The amplitude of this fourfold oscillation becomes very small when {\boldmath $H$} is rotated conically around the $c$-axis with a tilt angle of  45$^{\circ}$.    Based on these results, we  provide the first compelling evidence that the gap function of YNi$_2$B$_2$C has {\it point nodes}, which are located along the [100] and [010]-directions. This unprecedented gap structure challenges the current view on the pairing mechanism and on the unusual superconducting properties of borocarbide superconductors.

\end{abstract}

\maketitle

	   Superconductivity with unconventional pairing symmetry has been a central subject in the physics of superconductors.  The vast majority of superconductors have conventional $s$-wave symmetry with an isotropic gap which is independent of  directions over the entire Fermi surface. In the last two decades unconventional superconductivity with different symmetry has been found in heavy fermion materials \cite{sigrist}, high-$T_c$ cuprates\cite{tsuei}, ruthenate\cite{maeno}, and organic compounds \cite{kanoda}.   Unconventional superconductivity is characterized by anisotropic superconducting gap which becomes to zero (nodes) along certain crystal directions \cite{mineev}.  Since the nodal structure is ultimately related to the pairing interaction, its identification is crucial for understanding the pairing mechanism.   A common feature in the unconventional superconductors discovered until now is that they all have {\it line nodes} in the gap functions, which are located parallel or perpendicular to the basal planes. It is generally believed that these nodal structures appear as a result of  pairing interactions caused by purely electronic origin, instead of being mediated by conventional phonons.     In fact a close relation between the spin fluctuation and superconductivity has been discussed in all of these unconventional superconductors. 	

	  It is well known that there can be another type of gap singularity;   i.e. point nodes in which the gap turns to zero at isolated points on the Fermi surface.   The point node is thought to be present in the $A$-phase of superfluid $^3$He \cite{he3}.  On the other hand,  its existence has never been reported in any superconductor so far.  To our knowledge, the one exception is the $B$-phase of UPt$_3$ in which a possibility of  the point nodes coexisting with line nodes was suggested theoretically \cite{upt3},  but with no firm experimental evidence.   Thus it is still an open question whether a superconductor can have point nodes, and its clarification is an important subject for the study of the unconventional superconductivity.  

	 In this Letter,  we direct our attention to a new class of superconductors; non-magnetic borocarbide superconductors LnNi$_2$B$_2$C, Ln=(Y and Lu) \cite{cava}.  These borocarbide superconductors with elevated transition temperatures are in the clean regime $\xi \ll \ell$,where $\xi$ is the coherence length and $\ell$  the mean free path.   The electronic structure is essentially 3D and the crystal structure is tetragonal.  At an early stage, the gap symmetry of these materials was considered to be isotropic $s$-wave, mediated by conventional electron-phonon interactions \cite{eliashberg}.   However, recent experimental studies, such as specific heat \cite{nohara,izawa1}, thermal conductivity \cite{tail1},  Raman scattering \cite{raman}, and photoemission spectroscopy \cite{yokoya}  on  YNi$_2$B$_2$C or LuNi$_2$B$_2$C have reported a large anisotropic gap function.  Despite these studies, the detailed structure of the gap function remains unknown.   Here, we provide a strong evidence that the gap function of YNi$_2$B$_2$C has {\it point nodes},  which are located along the [100] and [010] directions.  These results are based on the angular variation of the thermal transport in a magnetic field rotated in various directions relative to the crystal axes, which has  proved  to be a powerful probe for determining the low energy quasiparticle (QP) excitation including its direction \cite{yu,aubin,izawa2,izawa3,vekhter1,maki1,fay,maki2,thalmeier}.   This unprecendented  gap structure challenges the current view on the mechanism of the superconductivity and on the many unusual properties in the superconducting state, both of which have been discussed without taking account the presence of the point nodes \cite{eliashberg,nonlocal}.  
		
	Single crystals of YNi$_2$B$_2$C  ($T_c$=15.5~K) were grown by the floating zone method.  The residual resistivity ratio was approximately 47, indicating the highest crystal quality  currently achievable.   We measured the $c$-axis thermal conductivity $\kappa_{zz}$ (the heat current {\boldmath $q$} $\parallel c$)  on the sample cut into a rectangular shape ($0.31\times0.54\times3.00$mm$^3$) by the steady-state method with one heater and two RuO$_{2}$ thermometers.    To apply {\boldmath $H$} with high accuracy relative to the crystal axes,  we used a system with two superconducting magnets generating {\boldmath $H$} in two mutually orthogonal directions and a $^{3}$He cryostat equipped on a mechanical rotating stage with a minimum step of 1/500 degree at the top of the Dewar \cite{izawa2,izawa3}.  Computer-controlling the two magnets and the rotating stage, we were able to rotate {\boldmath $H$} with a misalignment of less than 0.02$^{\circ}$  from the plane, which we confirmed by the simultaneous measurement of the resistivity $\rho_{zz}$. 

\begin{figure}
\includegraphics[scale=0.45]{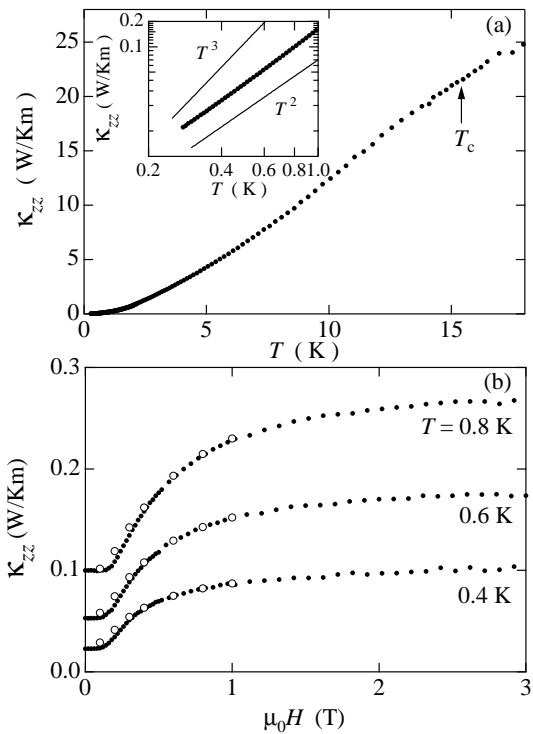}
\caption{(a)$T$-dependence of the $c$-axis thermal conductivity $\kappa_{zz}$ in zero field. Inset: Log-log plot of the same data below 1~K. (b)Field dependence of  $\kappa_{zz}$ at low temperatures ( {\boldmath $H$}$\parallel$ [110]).  The solid circles represent  the data measured by sweeping $H$ after zero field cooling and the open circles represent  the data measured under the field cooling conditions at each temperature. }
\end{figure}
	We first discuss the $T$- and $H$- dependence of $\kappa_{zz}$.     Figure 1 (a) shows the $T$-dependence of $\kappa_{zz}$ in zero field.   Upon entering the superconducting state, $\kappa_{zz}$ exhibits a tiny kink.   The Wiedemann-Franz ratio $L = \kappa_{zz}\rho_{zz}/T\simeq1.02L_0$ at $T_{c}$ is very close to the Lorenz number $L_{0}= 2.44\times 10^{-8}$~$\Omega$W/K, indicating that the electronic contribution well dominates over the phonon contribution.  The phonon peak at $\sim T_c / 3$  reported in LuNi$_2$B$_2$C  was not observed , possibly due to the difference in the phonon spectrum in the two compounds\cite{tail2}.   The inset of Fig.~1 (a) shows the same data  below 1~K.  The $T$-dependence of $\kappa_{zz}$ is close to quadratic rather than cubic.   We will discuss this $T$-dependence later.  Figure 1 (b) depicts the $H$-dependence of $\kappa_{zz}$ ({\boldmath $H$}$\parallel $[110]) at low temperaures.  At low field $\kappa_{zz}$ increases rapidly with $H$. This steep increase is markedly different from that observed in the ordinary $s$-wave superconductors in which the thermal conductivity shows an exponential behavior with  much slower growth with $H$ at $H \ll H_{c2}$.   It has been reported that in LuNi$_2$B$_2$C($T_c$=16.6~K), a close cousin of YNi$_2$B$_2$C,  the $H$-dependence of the in-plane thermal conductivity $\kappa_{xx}$ is even steeper than UPt$_3$ with line nodes \cite{tail1}.  The steep increase of the thermal conductivity in YNi$_2$B$_2$C and LuNi$_2$B$_2$C, along with the $\sqrt{H}$-dependence of the heat capacity \cite{nohara,izawa1}, are evidence that the thermal properties are governed by the delocalized QPs arising from the nodes in the gap function.  The theoretical understanding of the heat transport for superconductors with nodes has largely progressed during past few years \cite{vekhter2}.  The most remarkable effect on the thermal transport is the Doppler shift of the energy of a QP with momentum {\boldmath $p$} ($\varepsilon(\mbox{\boldmath $p$})\rightarrow \varepsilon(\mbox{\boldmath $p$})-\mbox{\boldmath $v$}_s \cdot \mbox{\boldmath $p$}$) in the circulating supercurrent flow $\mbox{\boldmath $v$}_s$ \cite{volovik}.  This effect becomes important at such positions where the local energy gap becomes smaller than the Doppler shift term ($\Delta < \mbox{\boldmath $v$}_s \cdot \mbox{\boldmath $p$}$), which can be realized in the case of superconductors with nodes.  This effect gives rise to a steep increase of the thermal conductivity with $H$, which is consistent with the present results.  		
		Having established the predominant contribution of the extended QPs in the thermal transport, the next question is the nodal structure of the gap function.  The important advantage of choosing to measure the thermal conductivity is that it is  a {\it directional} probe, sensitive to the relative orientation between the magnetic field and nodal directions of the order parameter \cite{yu,aubin,izawa2,izawa3,vekhter1,maki1,fay,maki2,thalmeier}.   This statement is based on the fact that the magnitude of the Doppler shift depends on the angle between the node and  {\boldmath $H$}.  For instance, when  {\boldmath $H$} is rotated within the basal plane in $d$-wave superconductors,  the Doppler shift gives rise to the fourfold oscillation of the density of states (DOS).  In this case, the DOS shows the maximum (minimum) when {\boldmath $H$} is applied to the antinodal (nodal) directions.  In fact, a clear fourfold modulation of the thermal conductivity, which reflects the angular position of line nodes of $d$-wave symmetry, has been observed in high-$T_c$ cuprate YBa$_2$Cu$_3$O$_{7-\delta}$ \cite{yu,aubin}, heavy fermion CeCoIn$_5$\cite{izawa2}, organic $\kappa$-(ET)$_2$Cu(NCS)$_2$ \cite{izawa3},  demonstrating that the thermal conductivity can be a relevant probe of the superconducting gap structure.  	
		
\begin{figure}
\includegraphics[scale=0.45]{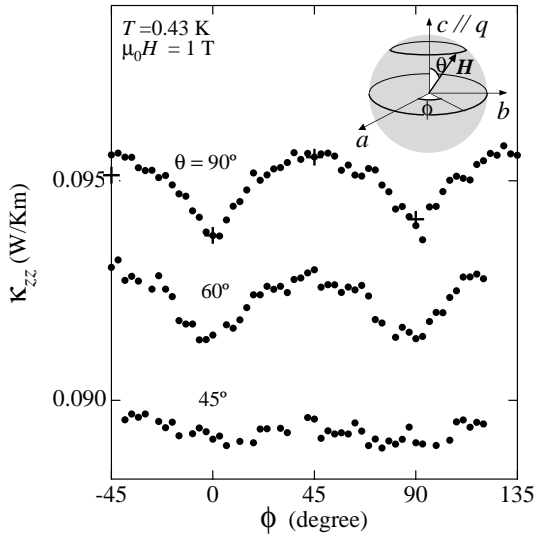}
\caption{Angular variation of the $c$-axis thermal conductivity $\kappa_{zz}$ at $H$=1~T and $T$=0.43~K  (thermal current {\boldmath $q$}$\parallel c$).  $\kappa_{zz}$ are measured by rotating   {\boldmath $H$}=$H(\sin\theta\cos\phi, \sin\theta\sin\phi, \cos\theta)$ conically as a function of $\phi$ at  fixed $\theta (=90^{\circ}$, 60$^{\circ}$, and 45$^{\circ}$).   $\theta$=({\boldmath $q$}, {\boldmath $H$}) is the polar angle and $\phi$ is the azimuthal angle measured from the $a$-axis (see the inset).   The crosses represent the data at  $\theta=90^{\circ}$ which are obtained under the field cooling condition at each angle. }
\end{figure}
		Figure 2 displays the angular variation of $\kappa_{zz}$, which was measured by rotating {\boldmath $H$}=$H(\sin\theta\cos\phi, \sin\theta\sin\phi, \cos\theta)$ conically as a function of $\phi$, keeping $\theta$ constant.  Here $\theta$=({\boldmath $q$}, {\boldmath $H$}) is the polar angle and $\phi$ is the azimuthal angle measured from the $a$-axis (see the inset of Fig.~2).    The measurements have been done by rotating $\phi$ after field cooling at $\phi=-45^{\circ}$.    The crosses in Fig.2 show $\kappa_{zz}(H, \phi)$ at $H$=1~T which are obtained under the field cooling condition at each angle.   $\kappa_{zz}(H,\phi)$ obtained by different procedures of field cooling well coincide with each other, indicating that the field trapping effect related to the vortex pinning is negligibly small at $H$=1~T.  Below 0.5~T,  a small field trapping effect was observed (see also Fig.~1(b)).  A clear fourfold symmetry is observed at $\theta=90^{\circ}$ and $60^{\circ}$; $\kappa_{zz}$ can be decomposed as  $\kappa_{zz}=\kappa_{zz}^0+\kappa_{zz}^{4\phi}$ where $\kappa_{zz}^0$ is a $\phi$-independent term and $\kappa_{zz}^{4\phi}$ is a term with fourfold symmetry with respect to $\phi$-rotation.  Similar results were obtained at 0.27~K and 0.8~K.  We note that in the previous measurements in which the in-plane thermal conductivity $\kappa_{xx}$ was measured in {\boldmath $H$} rotating within the $ab$-plane, a large twofold term appears as a result of the difference of the effective DOS for QPs traveling parallel to the vortex and for those moving in the perpendicular direction  \cite{yu,aubin,izawa2,izawa3}.  In the present geometry, on the other hand, such a twofold term is absent because out-of-plane thermal conductivity $\kappa_{zz}$ was measured in keeping $\theta$, i.e. the angle between  {\boldmath $H$} and  {\boldmath $q$}, constant.  This geometry enables us to make a precise analysis of the term with fourfold symmetry, which is directly related to the electronic structure \cite{fay}. 
	  
	  As seen in Fig.~2, $\kappa_{zz}^{4\phi}$ shows a peculiar angular variation.  The first point to emphasize is the appearance of a a narrow cusp at $\phi=0^{\circ}$ and 90$^{\circ}$.    This angular variation is markedly different from those in $d$-wave superconductors with line nodes, in which the fourfold oscillation is close to a sinusoidal wave \cite{yu,aubin,izawa2,izawa3}.   The second point is that the amplitude of $\kappa_{zz}^{4\phi}$ decreases rapidly as {\boldmath $H$} is changed from $\theta=90^{\circ}$ to $45^{\circ}$.    Here we stress that the fourfold anisotropies of the Fermi velocity $v_F$ and $H_{c2}$, which are inherent to the tetragonal band structure of YNi$_2$B$_2$C, are quite unlikely to be an origin of the observed fourfold symmetry for the following reasons.  We measured  the $\phi$-dependence of $H_{c2}$ at $\theta=90^{\circ}$ and 45$^{\circ}$ and  found that $H_{c2}$  shows nearly perfect sinusoidal $\phi$-dependence with fourfold symmetry, similar to $H_{c2}$ of LuNi$_2$B$_2$C reported in Ref.\cite{hc2}.   This is totally different from the $\phi$-dependence of $\kappa_{zz}^{4\phi}$ displayed in Fig.~2.   In addition,  the amplitude of fourfold oscillation of $H_{c2}$ at $\theta=45^{\circ}$  was nearly 1/3 of that at $\theta=90^{\circ}$, while the amplitude of $\kappa_{zz}^{4\phi}$ at $\theta=45^{\circ}$ is less than 1/5 of that at $\theta=90^{\circ}$, indicating a different $\theta$-dependence of two quantities.   These results are consistent with the theory of the thermal conductivity.  According to the calculation based on the Kubo formula, the leading term is the anisotropy of the gap function and anisotropy of $v_F$  will only enter as a secondary effect.  Thus these consideration lead us to conclude that {\it the observed fourfold symmetry originates from the superconducting gap nodes.}  

\begin{figure}
\includegraphics[scale=0.45]{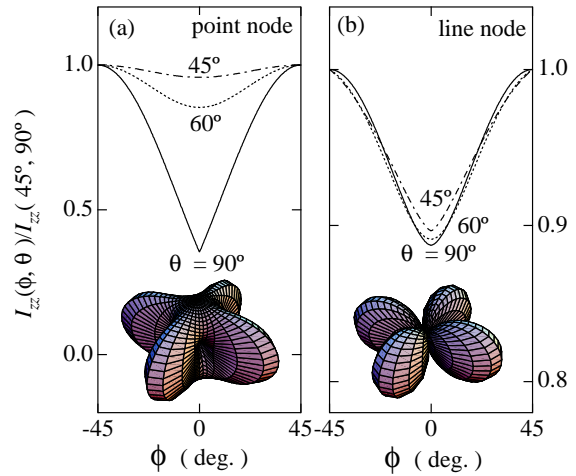}
\caption{Angular variation of $I_{zz}(\phi,\theta)$ normalized by $I_{zz}(45^{\circ},90^{\circ})$ as a function of $\phi$ at $\theta=90^{\circ}$ (solid lines), $60^{\circ}$ (dashed lines), and  $45^{\circ}$ (dash-dotted lines),  which are calculated by the gap functions with  (a) point node and (b)line node.  For the definition of $\theta$ and $\phi$, see the inset of Fig~2.   The corresponding gap functions are illustrated in the insets.   For details, see the text. }
\end{figure}
	We are now in position to discuss the nodal structure of YNi$_2$B$_2$C.  The fact that $\kappa_{zz}^{4\phi}$ shows the minimum at  $\phi=0^{\circ}$ and 90$^{\circ}$ immediately indicates that {\it the nodes are located along [100] and [010]-directions.}   The cusp structure and the $\theta$-dependence of $\kappa_{zz}^{4\phi}$ are key features for specifying the type of nodes.   	In Figs. 3 (a) and (b), we compare $\kappa_{zz}^{4\phi}$ for two different types of nodes calculated from the Doppler-shifted QP spectrum.  Here we adopted  gap functions $\Delta(\mbox{\boldmath $k$})=\frac{1}{2}\Delta_0 \{1-\sin^4\theta\cos(4\phi)\}$ ($s+g$-wave) proposed in Ref.\cite{maki2,thalmeier} for point node, and $\Delta(\mbox{\boldmath $k$})=\Delta_0\sin(2\phi)$ ($d$-wave)  for the line node.  These gap functions are illustrated in the insets of Figs.3 (a) and (b).   We caution readers that the  the gap functions are rotated by $45^{\circ}$ from those  of Ref.\cite{maki2,thalmeier}, since the nodes in these references are located along the [110] direction.   In {\boldmath $H$} for general orientation given by $(\theta,\phi)$, $\kappa_{zz}(H,\phi,\theta)$ is given as 
\begin{equation}
\frac{\kappa_{zz}}{\kappa_n}=\frac{1}{6\sqrt{2} \pi} 
\left( \frac{v_av_c\hbar eH}{\Delta^2 c} \right)^{\frac{3}{2}}
 I_{zz}(\phi,\theta)
\end{equation}
where $v_{a,c}$ are the anisotropic Fermi velocities.   For point nodes $I_{zz}$ is given as $I_{zz}(\phi,\theta)=I_{+}(\phi,\theta)^3$ and
\begin{eqnarray}
I_{+}(\phi,\theta)=\frac{1}{2}[\{1+\cos^2\theta+\sin^2\theta\cos(2\phi)\}^{\frac{1}{2}} \nonumber \\ 
+\{1+\cos^2\theta-\sin^2\theta\cos(2\phi)\}^{\frac{1}{2}}].
\end{eqnarray}
For the line nodes  $I_{zz}$ is given as $I_{zz}(\phi,\theta)=I_{+}(\phi,\theta)^2$ and 
\begin{eqnarray}
I_{+}(\phi,\theta)=\frac{1}{2\pi} \int^{\frac{\pi}{2}}_{-\frac{\pi}{2}}d\psi
[\{1+\frac{1}{2}\sin^2\theta (\cos(2\phi)-\cos(2\psi))\nonumber \\ +
\frac{1}{\sqrt{2}}\sin(2\theta)\sin\psi \sqrt{1-\sin(2\theta)}\}^{\frac{1}{2}} \nonumber \\ +\{1-\frac{1}{2}\sin^2\theta(\cos(2\phi)+\cos(2\psi)) \nonumber \\ + \frac{1}{\sqrt{2}}\sin(2\theta)\sin\psi\sqrt{1+\sin(2\theta)}\}^{\frac{1}{2}}].
\end{eqnarray}
Here, the superclean limit $\frac{\hbar\Gamma}{\Delta}\ll\frac{H}{H_{c2}}$ is assumed, where $\Gamma$ is the carrier scattering rate which is estimated to be $\hbar\Gamma<$1~K \cite{terashima}.  Using $\Delta\simeq$28~K, $H_{c2}\simeq$10~T, the superclean condition is well satisfied in the present measurement.    We first compare $\kappa_{zz}^{4\phi}$ at $\theta=90^{\circ}$.  A clear narrow cusp  structure is seen at $\phi=0^{\circ}$ and $90^{\circ}$ for the point node, while it is absent and $\kappa_{zz}^{4\phi}$ is close to the sinusoidal wave for the line node.    Qualitatively this can be explained as the following.  The narrow cusp  in $\kappa_{zz}^{4\phi}$  appears as a result of the smaller freedom of the QPs induced in the vicinity of point nodes.  On the other hand, the cusp structure is smeared out due to the larger freedom associated with the line node.    We next discuss how the amplitude of $\kappa_{zz}^{4\phi}$ changes with $\theta$.   For point nodes,  the amplitude of $\kappa_{zz}^{4\phi}$ at $\theta=45^{\circ}$ is much smaller than that at $\theta=90^{\circ}$, while they are of almost the same magnitude for line nodes.  This can be accounted for considering the fact that  rotating {\boldmath $H$} conically at $\theta=45^{\circ}$ does not point to the nodes in case of the point nodes, while rotating {\boldmath $H$} at any $\theta$ always cross the nodes in case of the line nodes.   It is apparent that both the cusp structure and $\theta$-dependence of $\kappa_{zz}^{4\phi}$ shown in Fig.~3(a) are strongly in favor of the point nodes.  We have calculated  $\kappa_{zz}^{4\phi}$ for different types of gap functions with point nodes and found that both the cusp structure and  the $\theta$-dependence of $\kappa_{zz}^{4\phi}$ are robust for the choice of the gap functions.  In addition, we have measured  the in-plane thermal conductivity  ({\boldmath $q$$\parallel $} [110] ) of a different crystal in {\boldmath $H$} rotated within the [110]-plane and found no anomaly associated with the point node along the [001]-direction.     These results lead us to our forementioned conclusion that  the superconducting gap function of YNi$_2$B$_2$C has {\it point nodes} which are located along [100] and [010]-directions. 
	
	For point nodes,  the $T$-dependence of the thermodynamic quantities at low temperatures strongly depend on the dispersion of the gap function near the nodes.   We point out that the gap function we used  predicts a  quadratic $T$-dependence of the thermal conductivity, which is consistent with the data in the inset of Fig.~1(a).   On the other hand, a cubic $T$-dependence of $\kappa_{xx}$ has been  reported in LuNi$_2$B$_2$C below 150~mK \cite{tail1,tail2}.   One possible origin for the discrepancy between YNi$_2$B$_2$C and LuNi$_2$B$_2$C is the difference of the dispersion of the gap.	
	
	We here comment on the gap anisotropy and the impurity effect.  In the present experiment, we cannot rule out the possibility that there remains a very small but finite gap at the nodal positions.   Let us estimate the upper limit of the gap anisotropy ratio $\Delta_{min}/\Delta_{max}$, where $\Delta_{max}$ and $\Delta_{min}$ are the gap maximum and minimum, respectively.  The fact that the clear fourfold pattern characterisitic to the point node was observed even at T=0.27~K suggests that $\Delta_{min}$ is much less than 0.27~K.  Assuming the weak coupling BCS relation for the maximum gap ($\Delta_{max}$=1.75$k_BT_c$), we estimate $\Delta_{max}\simeq$28~K.   We then obtain $\Delta_{min}/\Delta_{max}\ll 0.01$; the anisotropy ratio is essentially zero.     According to Ref.\cite{nohara,yokoya}, the introduction of non-magnetic impurity in  YNi$_2$B$_2$C reduces the DOS at the Fermi surface by removing the node.  This indicates that the gap function does not change its sign at the nodes,  because otherwise the impurity always induces a finite DOS at the Fermi level, as observed in high $T_c$ cuprates \cite{hirsch}.   Therefore the impurity effect is consistent with the point node for which case the gap function does not change its sign at the node.  
	 	 
	 Various authors have pointed out that  calculations based on the Eliashberg equation reproduces well the high transition temperature of borocarbide superconductors in terms of the observed phonon spectrum and isotropic gap \cite{eliashberg}.   However the determined gap structure with point nodes casts doubt on such view based on a simple phonon mediated pairing mechanism.  At the present stage,  what kind of the pairing interaction gives rise to the point node is not known.   We also point out that the determined nodal structure should play an important role in determining the superconducting properties, such as the vortex lattice structure, reversible magnetization, upper critical field $H_{c2}$, etc.   Until now, these preperties have been discussed in terms of an ordinary $s$-wave superconductor with anisotropic Fermi surface with the use of nonlocal London theory \cite{nonlocal}.   However,  the determined nodal structure motivates further investigations on these subjects.  For instance,  the extended QPs appear to be very important for the vortex triangular-square lattice phase transition \cite{vl}.   	 
	 
	 In summary,  we demonstrated that the measurement of  $\kappa_{zz}$ in {\boldmath $H$} rotating conically around the $c$-axis with a fixed tilt angle provides a novel method to distinguish line and point nodes.  We presented the first compelling evidence that the gap function of YNi$_2$B$_2$C has {\it point nodes}  which are located along the [100] and [010] directions.   To the best of our knowledge,  YNi$_2$B$_2$C ( and presumably LuNi$_2$B$_2$C) is the first superconductor in which the gap function with the point nodes was successufully identified.  The determined gaps structure offers a new perspective on the pairing mechanism as well as unusual superconducting properties of anisotropic superconductors.  

	We thank D.F.~Agterberg, T.~Dahm,  V.~Kogan,  K.~Machida, P.~Miranovic,  M.~Sigrist, A.~Tanaka, K.~Ueda and  I.~Vekhter for stimulating discussions.
	

%
\end{document}